\documentclass[aps,prb,twocolumn,showpacs,superscriptaddress]{revtex4-1}
\usepackage{amsfonts,amsmath,amssymb}
\usepackage[colorlinks=true,linkcolor=blue,pagecolor=blue,filecolor=blue,menucolor=blue,urlcolor=blue,citecolor=blue,anchorcolor=blue]{hyperref}%
\usepackage[dvips]{graphics}
\usepackage{graphicx}
\usepackage{bm}
\usepackage{color}

\begin{document}

\title{Quasiclassical theory of magnetoelectric effects in superconducting heterostructures in the presence of the spin-orbit coupling}
%\title{paper}
\author{I. V. Bobkova}
\affiliation{Institute of Solid State Physics, Chernogolovka, Moscow
  reg., 142432 Russia}
\affiliation{Moscow Institute of Physics and Technology, Dolgoprudny, 141700 Russia}
\author{A. M. Bobkov}
\affiliation{Institute of Solid State Physics, Chernogolovka, Moscow reg., 142432 Russia}

\date{\today}

\begin{abstract}
The quasiclassical theory in terms of equations for the Green's functions (Eilenberger equations) is generalized in order to allow for quantitative description of the magneto-electric effects and proximity-induced triplet correlations in the presence of spin-orbit coupling in hybrid superconducting systems. The formalism is valid under the condition that the spin-orbit coupling is weak with respect to the Fermi energy, but exceeds the superconducting energy scale considerably.
On the basis of the derived formalism it is shown that the triplet correlations in the spin-orbit coupled normal metal can be induced by proximity to a singlet superconductor without any exchange or external magnetic field. They contain an odd-frequency even-momentum component, which is stable against disorder. The value of the proximity-induced triplet correlations is of the order of $\Delta_{so}/\varepsilon_F$, that is absent in the framework of the standard quasiclassical approximation, but can be described by our theory.
The spin polarization, induced by the Josephson current flowing through the superconductor/Rashba metal/superconductor junction, is also calculated.

\end{abstract}

\maketitle

\section{introduction}

\label{introduction}

By now it is already known that spin-orbit coupling (SOC) is a source of many interesting phenomena. Some of them originate from coupling of charge and spin degrees of freedom and often called by magnetoelectric effects. In addition to their fundamental importance these phenomena can be of interest for the spintronics and, in particular, for the superconducting spintronics\cite{linder15,eschrig15}.
It is worth to mention here some of them, which are related to the subject of this work.  For nonsuperconducting systems these are the spin Hall effect (SHE)\cite{dyakonov71,dyakonov71_2,chazalviel75,hirsch99,dyakonov06,mishchenko04,kato04,kato04_2,wunderlich05,raimondi06}, the inverse SHE\cite{valenzuela06,morota11,isasa14}, the direct magneto-electric \cite{aronov89,edelstein90,kato04,silov04} and inverse magneto-electric (spin-galvanic) effects\cite{shen14,ganichev02,rojas_sanchez13}.

The direct magneto-electric effect was also predicted for superconducting systems \cite{edelstein95,edelstein05,malshukov08,bergeret16}, where it consists in generation of an equilibrium spin polarization in response to a supercurrent. The analogue of the inverse magneto-electric effect has also been reported for superconducting systems. For homogeneous superconducting systems its physics is that the SO-coupled superconductor turns into the inhomogeneous phase-modulated state (it is also called by the helical phase) in response to an applied exchange field \cite{edelstein89,samokhin04,kaur05,dimitrova07,houzet15}. For Josephson junction the inverse magneto-electric effect is a cause of the anomalous phase shift $\varphi_0$, which modifies the current-phase relation according to $j=j_c \sin (\varphi-\varphi_0)$. This is the so called $\varphi_0$-junction, and its interpretation in terms of the inverse magneto-electric effect was reported in \cite{konschelle15}. It was actively studied recently in half-metal junctions, noncoplanar ferromagnetic junctions, ferromagnetic Josephson junctions with spin orbit interaction or TI surface states \cite{krive04,braude07,asano07,reinoso08,buzdin08,tanaka09,zazunov09,liu10,alidoust13,brunetti13,yokoyama14,bergeret15,dolcini15,campagnano15,mironov15,konschelle15,kuzmanovski16,zyuzin16}.

The other group of effects, which are of entirely superconducting nature, is connected to the generation of triplet superconducting correlations
via proximity to a conventional superconductor (S). This singlet-triplet conversion can also be viewed as a coupling between charge and spin degrees of freedom, that is a kind of magnetoelectric effect.  The
triplet Cooper pairs with non-zero average spin can play the
same role in superconducting spintronics as electron spins in conventional spintronics. However, the corresponding spin currents are dissipationless.
Most of the research in this area, both experimental and
theoretical, has focused on proximity junctions involving ferromagnets (F) (see Ref.~\onlinecite{eschrig15} and references therein).
The ferromagnets induce triplet correlations
by lifting the spin degeneracy \cite{buzdin05,bergeret05}. But, the proximity between a singlet superconductor and a homogeneous ferromagnet
only leads to creation of zero average spin pairs. In order  to have non-zero average spin one needs to include
a magnetic inhomogeneity or SOC into consideration because they act as spin mixers \cite{bergeret05,eschrig15,bergeret13,bergeret14}.

The SOC also lifts spin degeneracy. It has been demonstrated that for homogeneous superconductors in the presence of SOC the pair
wave function is the mixture of singlet s-wave and triplet p-wave components \cite{gorkov01,alicea10}. So, the question is if
it by itself can induce triplet pair correlations in proximity to a singlet superconductor?
The most convenient and commonly used method to treat superconducting hybrid systems is a quasiclassical theory of superconductivity. The SOC can be treated in the quasiclassical approximation when its characteristic energy $\Delta_{so}$ is much less than the Fermi energy $\varepsilon_F$. This situation is typical. Within the framework of the quasiclassical theory it was found that SOC by itself does not induce
any triplet pairing \cite{bergeret13,bergeret14}.

On the other hand, working in the framework of Gor'kov equations beyond the quasiclassical approximation, Edelstein showed that interfacial
spin-orbit scattering generates triplet pairing in 3D ballistic superconductor/normal-metal junctions\cite{edelstein03}. More recently, it was reported in several works on the basis of the lattice numerical calculation \cite{yang09}, a gauge-covariant analytical approach \cite{konschelle15}, and, at last, on the basis of exact Gor'kov technique, that triplet superconductivity can be generated in Rashba metals by proximity to a singlet superconductor\cite{reeg15}.

However, the Gor'kov equations are of very limited use for inhomogeneous problems. So, it is desirable to generalize the quasiclassical technique in order to be able to describe the SOC-induced triplet correlations and the magneto-electric and spin-galvanic effect, which are also beyond the framework of quasiclassical approximation. A way to such a generalization has already been proposed in the framework of gauge-covariant Green's functions approach \cite{konschelle15}. But the results for the proximity-induced triplet correlations in the ballistic limit seem to be not fully coinciding with the results of exact Gor'kov's approach \cite{reeg15}. Also the appropriate normalization condition and boundary conditions for the quasiclassical Green's functions were not considered in Ref.~\onlinecite{konschelle15}.

It is also worth to note here that the quasiclassical formalism in the framework of the gauge-covariant approach has been developed also for nonsuperconducting systems \cite{gorini10,raimondi12}. On the basis of this formalism a generalized Boltzmann
equation for the charge and spin distribution functions was formulated. Then it was applied, in particular, to the investigation of the spin Hall and the inverse spin galvanic (Edelstein) effects.

In this work we generalize the quasiclassical equations, the normalization condition and the corresponding boundary conditions for the absolutely transparent interface in order to be able to calculate the Green's functions up to the first order with respect to the parameter $\Delta_{so}/\varepsilon_F$. This allows us to describe the magneto-electric effects and proximity-induced triplet correlation in the presence of SOC, while the quasiclassical approximation only provides Green's functions up to zero order with respect to this parameter and is not able to catch them. In order to check our formalism we consider the proximity effect at the interface between the the singlet superconductor and the Rashba metal in the ballistic limit. The result of Ref.~\onlinecite{reeg15} is recovered, if the exact expression of this work is properly expanded up to the first order in $\Delta_{so}/\varepsilon_F$. The statement of this work that the triplet correlations are absent in the first order with respect to this parameter is incorrect.

We also consider the direct magneto-electric effect in superconductor/Rashba metal/superconductor ballistic junction. Its essence is a creation of a stationary spin polarization in response to a Josephson electric current flowing through the junction. As far as we know, this effect has not been quantitatively calculated so far, while the direct magneto-electric effect in homogeneous Rashba superconductors in ballistic\cite{edelstein95} and diffusive\cite{edelstein05} systems were considered, and  the direct magneto-electric effect in superconductor/Rashba metal/superconductor diffusive junction was calculated as well\cite{malshukov08}.

The paper is organized as follows. In sections \ref{equations} and \ref{bc} we derive the quasiclassical formalism, which accounts for the corrections up to the first order with respect to the parameter $\Delta_{so}/\varepsilon_F$. In sec.~\ref{equations} the corresponding equations for the quasiclassical Green's function and the normalization condition are derived, while sec.~\ref{bc} is devoted to the derivation of the boundary conditions. The proximity-induced triplet correlations at Rashba metal/superconductor interface are considered in sec.~\ref{triplet}, and the direct magneto-electric effect in superconductor/Rashba metal/superconductor ballistic junction is calculated in sec.~\ref{direct}. In sec.~\ref{conclusions} we summarize our results. The Appendix A is devoted to details of the direct magneto-electric effect calculations.

\section{generalized quasiclassical equations}

\label{equations}

In this section we derive the equations of motion for the quasiclassical Green's functions keeping the terms up to the first order with respect to the parameter $\Delta_{so}/\varepsilon_F$, while the standard quasiclassical approximation neglects them. It is these terms that provide singlet-triplet conversion in superconducting proximity systems in the absence of an applied magnetic field and/or any ferromagnetic elements. They are also responsible for the magneto-electric effects in SO-coupled systems.

We do not restrict ourselves by equilibrium situations and work with the Green's functions in the Keldysh technique \cite{keldysh,mahan}. We start with the Hamiltonian of a singlet superconductor in the presence of a generic linear in momentum spin-orbit (SO) coupling \cite{bergeret13,bergeret14}:

\begin{eqnarray}
\hat H=\int d^2 \bm r' \hat \Psi^\dagger (\bm r')\hat H_0(\bm r')\hat \Psi(\bm r')+  \nonumber \\
\Delta(\bm r)\Psi_\uparrow^\dagger (\bm r)\Psi_\downarrow^\dagger (\bm r)+\Delta^*(\bm r)\Psi_\downarrow (\bm r)\Psi_\uparrow (\bm r)
\label{H},
\end{eqnarray}
\begin{equation}
\hat H_0(\bm r)=\frac{\hat {\bm p}^2}{2m}-\frac{1}{2}\hat {\bm A} \hat {\bm p} - \hat h(\bm r) + V_{imp}(\bm r)-\mu
\label{H0},
\end{equation}
where $\Delta(\bm r)$ is the superconducting parameter and $\hat H_0$ is the Hamiltonian of the normal metal in the presence of the spin-orbit coupling (NSO). The general linear in momentum SO is expressed by the term $\frac{1}{2}\hat {\bm A} \hat {\bm p}=\frac{1}{2} A_{j}^\alpha p_j\hat \sigma^\alpha$, where  $\hat \sigma^\alpha$ are Pauli matrices in spin space. $\hat \Psi=(\Psi_\uparrow, \Psi_\downarrow)^T$, $\mu$ is the chemical potential, and
$\hat h=h^\alpha \hat \sigma^\alpha$ is an exchange field. We assume
that the system involves nonmagnetic impurities that can be described
by a Gaussian scattering potential: $V_{imp}(\bm r)=\sum \limits_{\bm r_i} V_i \delta(\bm r - \bm r_i)$.

The advanced ($A$), retarded ($R$), and Keldysh ($K$) blocks of Gor'kov Green function $\check G(\bm r_1, \bm r_2, t_1,t_2)$
in the Keldysh technique are defined in a standard way (see, for example, Ref.~\onlinecite{bbza16}).

By averaging the Green function over the impurity scattering potential in the Born approximation we find the following Gor'kov equation:
\begin{eqnarray}
\left[
i{\partial_{t_1}}\hat \tau_z-\frac{\hat {\bm p}^2}{2m}+\frac{1}{2}\hat A_j \hat p_j+\hat h(\bm r)\tau_z+\mu+\check \Delta-\check \Sigma
\right]\check G= \nonumber \\
\delta(\bm r_1-\bm r_2)\delta(t_1-t_2)
\label{Gor'kov_gen}.~~~~~~~~~
\end{eqnarray}
Here, we introduce the Pauli matrices in particle-hole space $\hat \tau_i $, with $\hat \tau_{\pm}=(\hat \tau_x \pm i \hat \tau_y)/2$.
$\check \Delta=\Delta \hat \tau_+ - \Delta^* \hat \tau_- $ is the matrix structure of the superconducting order parameter in the particle-hole space.
$\check \Sigma (\bm r_1)=\frac{1}{\pi N_F \tau}\check G(\bm r_1,\bm r_1)$ is the self-energy describing the elastic scattering at nonmagnetic impurities, where $\tau$ is the quasiparticle mean free time and $N_F$ is the density of states at the Fermi level of the normal state. The two time dependent products of  $AB$ operators is equivalent to $AB(t_1,t_2) \equiv \int dt' A(t_1,t')B(t',t_2)$.

In this work we concentrate on the ballistic systems, therefore below we assume $\tau \to \infty$. The including of the impurity self-energy into the resulting equations is straightforward.

The main goal of the present work is to develop the theory for plane interfaces between the SO materials and superconductors. We focus on the two-dimensional case here. For this reason it is convenient to perform the Fourier transformation with respect to the $y$-coordinate, parallel to the considered 2D interface:
\begin{equation}
\check G(\bm r_1,\bm r_2)=\int \frac{dp_y }{2\pi}\check G(p_y, y, x_1,x_2)e^{ip_y(y_1-y_2)},
\label{fourier_py}
\end{equation}
where $y=(y_1+y_2)/2$ and for generality we allow for the slow dependence of the Green's function on the center of mass coordinate $y$ along the interface. Substituting Eq.~(\ref{fourier_py}) into Eq.~(\ref{Gor'kov_gen}) one can obtain the Gor'kov equation for $\check G(p_y, y, x_1,x_2)$:
\begin{eqnarray}
\left[
i{\partial_{t_1}}\hat \tau_z-\frac{p_y^2}{2m}+\mu+\frac{ip_y}{2m}\partial_y+\frac{\partial_y^2}{8m}+\frac{\partial_{x_1}^2}{2m}+ \right.\nonumber \\
\frac{\hat A_y p_y}{2}-\frac{i}{4}\hat A_y \partial_y - \frac{i}{2}\hat A_x \partial_{x_1}+  \nonumber \\
\left.\hat h(\bm r)\tau_z+\check \Delta
\right]\check G=\delta(\bm x_1-\bm x_2)\delta(t_1-t_2)
\label{Gor'kov_py}.
\end{eqnarray}
Following the derivation of the quasiclassical equations presented in Refs.~\onlinecite{zaitsev84,millis88} we introduce the anzatz for the Gor'kov Green function:
\begin{eqnarray}
\check G(p_y, y, x_1,x_2)=\frac{1}{|v_{Fx}|}\left[
\check G_{11}e^{i|p_{F,x}|(x_1-x_2)}+ \times \right. \nonumber \\
\check G_{22}e^{-i|p_{Fx}|(x_1-x_2)}+\check G_{12}e^{i|p_{Fx}|(x_1+x_2)}+ \nonumber \\
\left. \check G_{21}e^{-i|p_{Fx}|(x_1+x_2)}\right],
\label{anzatz}
\end{eqnarray}
where the envelope functions $\check G_{ij}=\check G_{ij}(p_y, y, x_1, x_2)$ are slow functions of $(y, x_1, x_2)$, varying at quasiclassical length scales, except at $x_1=x_2$, where they are discontinuous.

Substituting anzatz (\ref{anzatz}) into Eq.~(\ref{Gor'kov_py}) we get for the envelope Green's functions (at $x_1 \neq x_2$) the following equation:
\begin{eqnarray}
\left[
i{\partial_{t_1}}\hat \tau_z+\hat h(\bm r_1)\tau_z+\check \Delta(\bm r_1)+\frac{1}{2}\hat A_y p_y-\right. \nonumber \\
\frac{1}{2}\hat A_x |p_{F,x}|(-1)^k-(-1)^k i|v_{F,x}|\partial_{x_1}+i\frac{v_y}{2}\partial_y+ \nonumber \\
\left.\frac{\partial_{x_1}^2}{2m}-\frac{i}{2}\hat A_x \partial_{x_1}-\frac{i}{4}\hat A_y \partial_y+\frac{\partial_y^2}{8m}\right]\check G_{kn}=0
\label{G_kn_1}.
\end{eqnarray}
The equivalent equation in the variable $x_2$ is
\begin{eqnarray}
\check G_{kn}\left[
-i{\partial_{t_2}}\hat \tau_z+\hat h(\bm r_2)\tau_z+\check \Delta(\bm r_2)+\frac{1}{2}\hat A_y p_y-\right. \nonumber \\
\frac{1}{2}\hat A_x |p_{F,x}|(-1)^n+(-1)^n i|v_{F,x}|\partial_{x_2}-i\frac{v_y}{2}\partial_y+ \nonumber \\
\left.\frac{\partial_{x_2}^2}{2m}+\frac{i}{2}\hat A_x \partial_{x_2}+\frac{i}{4}\hat A_y \partial_y+\frac{\partial_y^2}{8m}\right]=0
\label{G_kn_2}.
\end{eqnarray}
Further we reduce the amount of information by defining envelope functions of one variable,
\begin{eqnarray}
\check G_{kn}(x_1) \equiv \check G_{kn}(x_1,x_1+0),
\label{G_x}
\end{eqnarray}
which are closely related to the quasiclassical Green's functions (are defined below). Assuming $x_2=x_1+0=x+0$ in Eq.~(\ref{G_kn_1}), $x_1=x_2-0=x$ in Eq.~(\ref{G_kn_2}) and subtracting these equations, one obtains that $\check G_{kk}(x)$ obeys the following equation:
\begin{widetext}
\begin{eqnarray}
i{\partial_{t_1}}\hat \tau_z\check G_{kk} + i{\partial_{t_2}} \check G_{kk}\hat \tau_z + i {\bm v}_F \bm \nabla \check G_{kk} +
\left[ \check \Delta(\bm r)+\hat h(\bm r)\tau_z + \frac{1}{2}\hat A_y p_y-\frac{1}{2}\hat A_x |p_{F,x}|(-1)^k, \check G_{kk}\right]+~~~~~~ \nonumber \\
\left[ \frac{\partial_{x_1}^2}{2m}-\frac{i}{2}\hat A_x \partial_{x_1}-\frac{i}{4}\hat A_y \partial_y \right]\check G_{kk}^q(x_1,x_2)\Bigl |_{x_1=x_2-0=x}
-\check G_{kk}^q(x_1,x_2)\left[ \frac{\partial_{x_2}^2}{2m}+\frac{i}{2}\hat A_x \partial_{x_1}+\frac{i}{4}\hat A_y \partial_y \right]\Bigl |_{x_1=x_2-0=x} =0
\label{eq_G_x},~~~~~~
\end{eqnarray}
\end{widetext}
where ${\bm v}_F=((-1)^{k+1}|v_{F,x}|,v_y)$. Below we also use the analogous definition for ${\bm p}_F$. In Eq.~(\ref{eq_G_x}) the second line contains terms, which have additional small factor $(\Delta,h,\varepsilon, \Delta_{so})/\varepsilon_F$ with respect to the terms in the first line. Here $\Delta_{so}$ is the characteristic spin-orbit energy, which is of the order of $|A_i^j| p_F$. The terms in the first line represent the well-known quasiclassical Eilenberger equation \cite{eilenberger68,larkin69,bergeret14}, and the terms in the second line are corrections to the quasiclassical approximation and usually are neglected. However, as was already mentioned in the introduction, part of them are responsible for the magnetoelectric effects in superconductors and superconducting heterostructures and singlet-triplet conversion in the absence of the exchange field, therefore we should keep them in order to get possibility to treat these effects.

Further we will only keep the terms of the order of $\Delta_{so}/\varepsilon_F$, but will neglect the terms, which do not contain $\Delta_{so}$ (of the order of $(\Delta, \varepsilon, h)/\varepsilon_F$), because here we are interested in the limit $(\Delta,\varepsilon,h) \ll \Delta_{so} \ll \varepsilon_F$, which is appropriate for many real spin-orbit materials, such as metal surfaces\cite{lashell96,hoesch04,koroteev04} and
metallic surface alloys\cite{nakagawa07,ast07,eremeev12}.  Working in the framework of the perturbation theory up to the first order in $\Delta_{so}/\varepsilon_F$ it is enough to change the full Green's function $\check G$ by its quasiclassical approximation $\check G^q$. This allows us to simplify Eq.~(\ref{eq_G_x}) further. One can find $\partial_{x_{1,2}}\check G_{kk}^q(x_1,x_2)$ from the quasiclassical version of Eqs.~(\ref{G_kn_1}) and (\ref{G_kn_2}), respectively. For example, for $\partial_{x_{1}}\check G_{kk}^q(x_1,x_2)$ we get
\begin{eqnarray}
\partial_{x_1} \check G_{kk}^q(x_1,x_2) = \frac{(-1)^k}{i|v_{F,x}|}\left[ i{\partial_{t_1}}\hat \tau_z + \check \Delta(x_1)+\hat h(x_1)\tau_z + \right. \nonumber \\
\left.\frac{1}{2}\hat A_y p_y-\frac{1}{2}\hat A_x |p_{F,x}|(-1)^k +i \frac{v_y}{2}\partial_y \right]\check G_{kk}^q(x_1,x_2)
\label{x_derivative},~~~~~~
\end{eqnarray}
and an analogous expression can be found for $\partial_{x_{2}}\check G_{kk}^q(x_1,x_2)$ from Eq.~(\ref{G_kn_2}).
From Eq.~(\ref{x_derivative}) one can also obtain that
\begin{eqnarray}
\partial_{x_1}^2 \check G_{kk}^q(x_1,x_2) = -\frac{1}{v_{F,x}^2} \Biggl[\biggl\{ \left[ i{\partial_{t_1}}\hat \tau_z + \check \Delta(x_1)+ \right.  \nonumber \\
\left. \hat h(x_1)\tau_z+i \frac{v_y}{2}\partial_y \right],\frac{1}{2}\hat {\bm A} {\bm p}_F \biggr\}+
\frac{1}{4}(\hat {\bm A} {\bm p}_F)^2 \Biggr]\check G_{kk}^q
\label{x_derivative_2},~~~~~~
\end{eqnarray}
where $\left\{...,...\right\}$ means anticommutator, and we have neglected all the terms, which does not contain $\Delta_{so}$. For example, all the terms, proportional to spatial derivatives of $\Delta(x)$ and $\bm h(x)$ are disregarded. It is also assumed that the SO coupling $\hat {\bm A}$ does not depend on coordinates.

Substituting Eqs.~(\ref{x_derivative}), (\ref{x_derivative_2}) and the analogous expressions for the derivatives with respect to $x_2$ into Eq.~(\ref{eq_G_x}), we finally get for the envelope Green's functions:
\begin{widetext}
\begin{eqnarray}
i{\partial_{t_1}}\hat \tau_z\check G_{kk} + i{\partial_{t_2}} \check G_{kk}\hat \tau_z + i {\bm v}_F \bm \nabla \check G_{kk} +
\left[ \check \Delta(\bm r)+\hat h(\bm r)\tau_z + \frac{1}{2}\hat {\bm A} {\bm p}_F, \check G_{kk}\right]- \nonumber \\
\frac{1}{4v_{F,x}p_{F,x}}\Biggl( \biggl\{ \bigl[ \hat h \hat \tau_z, \hat A_x p_{F,x} \bigr],\check G_{kk}^q \biggr\}-p_{F,x}p_y \bigl( \hat A_x \hat A_y \check G_{kk}^q-\check G_{kk}^q \hat A_y \hat A_x \bigr)+i \bigl\{ \hat A_y {\bm v}_{F}{\bm p}_{F}, \partial_y \check G_{kk}^q \bigr\}+ \nonumber \\
\biggl\{ i{\partial_{t_1}}\hat \tau_z+\check \Delta + \hat h \hat \tau_z , \hat A_y p_y \biggr\}\check G_{kk}^q - \check G_{kk}^q \biggl\{ -i{\partial_{t_2}}\hat \tau_z+\check \Delta + \hat h \hat \tau_z , \hat A_y p_y \biggr\} \Biggr)=0
\label{eq_G_final},~~~~~~
\end{eqnarray}
\end{widetext}
where $p_{F,x}=(-1)^{k+1}|p_{F,x}|$.
The quasiclassical Green's functions are defined via the envelope functions as follows \cite{zaitsev84,millis88}:
\begin{eqnarray}
\check g_\pm = 2i \check G_{kk} +{\rm sgn}v_{F,x}\delta(t_1-t_2)
\label{g_def},
\end{eqnarray}
where the trajectory marked by the subscript "+"("-") is defined by $k=1(2)$ and $v_{F,x}>0(<0)$. Expressing the envelope functions via $\check g_\pm$ in Eq.~(\ref{eq_G_final}) it is easy to obtain the final equation for the quasiclassical Green's function. The second and third lines in Eq.~(\ref{eq_G_final}) represent the corrections of the first order in $\Delta_{so}/\varepsilon_F$ to the well-known quasiclassical equation, expressed by the first line of Eq.~(\ref{eq_G_final}).

Further we only consider  stationary problems, therefore a Fourier transformation with respect to $t_1-t_2 \to \varepsilon$ can be performed. We are interested in situations, when zeeman field is absent: $\hat h =0$. We also assume that the Green's function does not depend on the $y$-coordinate along the interfaces. Under these conditions Eq.~(\ref{eq_G_final}) can be simplified considerably and takes the form (it is already rewritten in terms of the quasiclassical Green's function):
\begin{eqnarray}
i {\bm v}_F \bm \nabla \check g +
\left[ \varepsilon \hat \tau_z + \check \Delta(\bm r) + \frac{1}{2}\hat {\bm A} {\bm p}_F, \check g\right]+ \nonumber \\
\frac{p_y}{4v_{F,x}} \bigl[ \hat A_x, \hat A_y \bigr]\bigl(\check g^q-{\rm sgn}v_{F,x} \bigr)+\frac{i\hat A_y p_y}{2p_{F,x}}\partial_x \check g^q=0
\label{eq_g_final}
\end{eqnarray}
Here we use the fact that $\left[ \check g^q,\hat A_i \right]=0$ at $\hat h=0$. This equation is one of the central results of our paper and contain all necessary terms to catch the proximity induced triplet correlations a NSO/S interface and the direct magnetoelectric effect in homogeneous superconductors \cite{edelstein95} and in ballistic superconducting heterostructures.

Eq.~(\ref{eq_g_final}) should be supplied by the normalization condition. In usual quasiclassical theory the normalization condition is $\check g^2=1$. However, we obtain that it should be modified if one would like to take into account the terms of the order of $\Delta_{so}/\varepsilon_F$. Below we derive the appropriate normalization condition. Multiplying Eq.~(\ref{G_kn_1})  by $\check G_{kk}$ from the left and Eq.~(\ref{G_kn_2}) by $\check G_{kn}$ from the right, we obtain the following expression at $\hat h=0$:
\begin{eqnarray}
\bigl(i v_{F,x} \partial_{x_1}+i \frac{v_y}{2}\partial_y \bigr) \bigl[\check G_{kk}(x_0,x_1)\check G_{kn}(x_1,x_2)\bigr]- \nonumber \\
\frac{i \bm p_F \bm v_f}{4 v_{F,x} p_{F,x}}\partial_y \bigl[ \check G_{kk}^q(x_0,x_1)\hat A_y \check G_{kn}^q(x_1,x_2) \bigr] = 0,
\label{norm_1}
\end{eqnarray}
where only terms of zero and first order in $\Delta_{so}/\varepsilon_F$ are kept, but all the terms of the first order with respect to $\Delta/\varepsilon_F$ are neglected. Please note that at $\hat h \neq 0$ Eq.~(\ref{norm_1}) is not valid. If the Green's functions do not depend on the $y$-coordinate, then the normalization condition for the envelope functions takes the well-known \cite{zaitsev84,millis88} form:
\begin{eqnarray}
\check G_{kk}(x_0,x_1)\check G_{kn}(x_1,x_2)=const
\label{norm_2}
\end{eqnarray}
for an arbitrary value of $x_1$. This $const$ can be easily found if the Eilenberger equation is solved for a half-space. For example, if we consider the left half-space, then one should take the case $x_{0,2}>x_1$. These inequalities cannot be changed because the envelope functions have discontinuities at coinciding arguments. Then taking the limit $x_0=x_2=x$ and $x_1 \to -\infty$ we get that $const=0$. Analogously for the right half-space one should take the case $x_{0,2}<x_1$. Then taking the limit $x_0=x_2=x$ and $x_1 \to +\infty$ we also get that $const=0$.

Further, the normalization condition for the quasiclassical Green's function can be obtained from Eq.~(\ref{norm_2}) at $x_0 = x_2 =x$ and $x_1=x \mp 0$ the sign $\mp$ corresponds to the left (right) half-space. The envelope function $\check G_{kk}(x,x+0)$ is directly connected to the quasiclassical Green's function $\check g(x)$ according to Eq.~(\ref{g_def}). In order to connect the envelope function $\check G_{kk}(x+0,x)$ to the quasiclassical Green's function, we need to calculate the discontinuity of $\check G_{kk}(x_1,x_2)$ at $x_1=x_2$. It can be obtained by integrating the Gor'kov equation (\ref{Gor'kov_py}) about $x_1 \approx x_2$ and taking into account the continuity condition for the full Gor'kov Green's function at $x_1=x_2$ \cite{zaitsev84,millis88}. Up to the first order terms with respect to $\Delta_{so}/\varepsilon_F$ we get the following expression:
\begin{eqnarray}
\check G_{kn}(x+0,x)=\check G_{kn}(x-0,x)- \nonumber \\
(-1)^{k+1}i\biggl(1 - \frac{ \hat A_y p_y}{2 v_{F,x}p_{F,x}}\biggr)\delta (t_1-t_2) \delta_{kn}
\label{discontinuity}
\end{eqnarray}
The second term in brackets represents the first order terms with respect to $\Delta_{so}/\varepsilon_F$. Substituting Eq.~(\ref{discontinuity}) together with Eq.~(\ref{g_def}) into Eq.~(\ref{norm_2}) at $x_0 = x_2 =x$ and $x_1=x \mp 0$ we get the following normalization condition:
\begin{eqnarray}
\check g^2 - \frac{\hat A_y p_y {\rm sgn}v_{F,x}}{p_{F,x}v_{F,x}}\biggl[ \check g^q - {\rm sgn}v_{F,x}\biggr]=1
\label{norm_final}
\end{eqnarray}
The same normalization condition is also valid for the regions, which are not semi-infinite, if $\check g^{q,2}=1$ is fulfilled there. It can be proven directly by multiplying Eq.~(\ref{eq_g_final}) by $\check g^q$ from the left, then from the right and adding the resulting equations.

\section{generalized boundary conditions}

\label{bc}

Quasiclassical equations are not valid in the vicinity of interfaces, where the normal state hamiltonian of the system changes over the atomic length scales. Therefore they should be supplied by the boundary conditions. In order to derive the appropriate boundary condition we generally follow Refs.~\onlinecite{zaitsev84,millis88}. The main strategy is to solve the interface scattering problem disregarding all the low-energy terms in the hamiltonians of the left and right materials: the superconducting order parameter, the quasiparticle energy and the exchange field should be neglected because we work only up to the zero order with respect to the parameter $(\Delta,h,\varepsilon)/\varepsilon_F$. In the standard quasiclassical approach the spin-orbit coupling term also should be neglected upon considering the scattering problem. However, our goal is to  correctly account for the terms of the first order with respect to $\Delta_{so}/\varepsilon_F$. Therefore, we must keep the spin-orbit coupling terms in the normal state hamiltonian of the SO-material.

Further we restrict ourselves by the case of "absolutely transparent interfaces" only. It means that where is no interface scattering barrier and there is no mismatch of the Fermi surfaces at the interface (without taking into account the spin-orbit coupling term), that is $\bm p_F^l = \bm p_F^r$. It is well-known that in this case the boundary conditions take the most simple linear form, while they are highly nonlinear for an arbitrary transparency of the interface and special further efforts are necessary to make them ready for practical use \cite{eschrig00,eschrig09,zhao04}. We postpone this problem for future consideration and demonstrate that even for the case of "absolutely transparent interfaces" the boundary conditions should be modified with respect to the standard form if we need to account for the terms of the first order with respect to $\Delta_{so}/\varepsilon_F$.

The Schrodinger equation in the interface region takes the form
\begin{eqnarray}
\hat H_0(\bm x)\check \Psi (x)=0,
\label{Schr}
\end{eqnarray}
where $\hat H_0(\bm x)=-(1/2m)\partial_x^2+p_y^2/2m-\mu - (1/2)\hat A_y (x) p_y + (i/2)\hat A_x (x) \partial_x $ and we assume $\hat A_{x,y}(x)=\hat A_{x,y}\Theta(-x)$, that is SO coupling is nonzero only on the left side of the interface. Its full solution can be written as follows:
\begin{eqnarray}
\check \Psi(x)^{l,r}=\frac{1}{\sqrt{|v_{F,x}|}}\sum \limits_{\alpha=1,2} \check \Psi_\alpha^{l,r}e^{ (-1)^{\alpha + 1}i|p_{F,x}|x},
\label{Schr_solution}
\end{eqnarray}
where at $x \ll |v_{F,x}|/\Delta_{so}$ ~$\check \Psi_\alpha^{l,r}$ are constant Nambu vectors, corresponding to left-moving ($\alpha=2$) and right-moving ($\alpha=1$) solutions. The general solution of the scattering problem at the interface between the SO-material and a material without SO-coupling can be easily found making use of Eq.~(\ref{Schr}) and the appropriate boundary conditions at the interface ($x=0$):
\begin{eqnarray}
\check \Psi(x) {\biggl |}_{x=-0}=\check \Psi(x) {\biggl |}_{x=+0},~~~~~~~~ \nonumber \\
\biggl[\frac{-i}{2m}\partial_x \check \Psi(x)-\frac{1}{4}\hat A_x \check \Psi(x)\biggr] {\biggl |}_{x=-0}=\frac{-i}{2m}\partial_x \check \Psi(x) {\biggl |}_{x=+0}
\label{Schr_bc}
\end{eqnarray}
The connection between the left and right-moving solutions  can be formulated in terms of the so-called interface transfer matrix $\hat M_{\alpha \beta}$ as follows
\begin{eqnarray}
\check \Psi_\alpha^{l}=\sum \limits_{\beta=1,2} \hat M_{\alpha \beta} \check \Psi_\beta^{r}.
\label{transfer_eq}
\end{eqnarray}
where for the considered here problem of the absolutely transparent interface
\begin{eqnarray}
\hat M = \left(
\begin{array}{cc}
\hat M_{11} & \hat M_{12} \\
\hat M_{21} & \hat M_{22} \\
\end{array}
\right)=
\left(
\begin{array}{cc}
1 \mp \hat {\delta S} & \pm \hat {\delta S} \\
\pm \hat {\delta S} & 1 \mp \hat {\delta S} \\
\end{array}
\right),
\label{M}
\end{eqnarray}
where upper and lower signs correspond to the NSO/S and S/NSO interfaces, respectively. $\hat {\delta S}=(\hat A_y p_y)/(4 v_{F,x}p_{F,x})$.

From Eq.~(\ref{transfer_eq}) and the conjugated equation one obtains that the envelope functions  $\check G_{kn}^l(x=-0)$ and $\check G_{kn}^r(x=+0)$ are connected by
\begin{eqnarray}
\check G_{\alpha \beta}^l(x=-0)=\sum \limits_{\mu,\nu=1,2} \hat M_{\alpha \mu} \check G_{\mu \nu}^r(x=+0) \hat M_{\nu \beta}^\dagger.
\label{transfer_eq_Green}
\end{eqnarray}
We are interested only in the boundary conditions for the envelope functions $\check G_{ii}(x)$ for the coinciding subscripts because only these envelope functions are connected to the quasiclassical Green's functions and are necessary for calculating observables. From Eq.~(\ref{transfer_eq_Green}) it follows that the boundary condition for $\check G_{11}(x)$ at NSO/S interface takes the form
\begin{eqnarray}
\check G_{11}^l= (1-\hat {\delta S})\check G_{11}^r (1-\hat {\delta S})+\check G_{12}^{r,q} \hat {\delta S}+\hat {\delta S}\check G_{21}^{r,q} ,
\label{boundary_G_11}
\end{eqnarray}
where we have taken into account that $\hat{\delta S}$ is of the first with respect to $\Delta_{so}/\varepsilon_F$, consequently all the terms, quadratic with respect to $\hat{\delta S}$ should be disregarded. For the same reason only the quasiclassical approximation for $\check G_{ij}(x)$ at $i \neq j$ enters the above equation. The boundary condition at S/NSO interface is obtained from Eq.~(\ref{boundary_G_11}) by substituting $\hat {\delta S} \to - \hat {\delta S}$.

It can be shown that $\check G_{12}^q (x)=\check G_{21}^q (x)=0$ in the ballistic limit and for the fully transparent interface we consider. In this case, taking into account the definition of the quasiclassical Green's function Eq.~(\ref{g_def}), one can obtain from Eq.~(\ref{transfer_eq_Green}) the following simple form of the boundary conditions
\begin{eqnarray}
\check g^l - \check g^r =\pm \Biggl\{ {\rm sgn}v_{F,x}-\check g^{q}, \frac{\hat A_y p_y}{4 v_{F,x}p_{F,x}} \Biggr\},
\label{boundary_final}
\end{eqnarray}
where the signs $\pm$ correspond to the NSO/S and S/NSO interfaces, respectively. It is seen that neglecting the right hand side of the above equation, which is of the first order in $\Delta_{so}/\varepsilon_F$, we obtain the well-known quasiclassical boundary condition at a fully transparent interface: $\check g^{l,q} = \check g^{r,q}=\check g^q$, that is just the continuity of the Green's function. This value of the quasiclassical value of the Green's function at the interface enters the right hand side of the boundary condition. It is worth to note here that if there is an equal SO coupling in the both materials, the boundary condition reduces to the standard continuity condition $\check g^l=\check g^r$.

\section{proximity-induced triplet correlations at NSO/S interface}

\label{triplet}

Here on the basis of the derived formalism we consider the proximity effect at a NSO/S interface, where the spin-orbit coupling in NSO is assumed to be of the Rashba-type $A_x^y=-A_y^x=\alpha$ for concreteness. It is found that taking into account the corrections of the first order with respect to $\Delta_{so}/\varepsilon_F$ to the quasiclassical approximation leads to the appearance of the proximity-induced triplet correlations in the NSO region without any exchange or Zeeman term. These correlations are long-ranged, that is they decay on the length scale of the normal state coherence length in the NSO region. They also contain an odd-frequency even-momentum component, which does not disappear after averaging over trajectories. These results are in sharp contrast with the results of the pure quasiclassical approximation, where the spin-orbit interaction by itself cannot be a source of any triplet correlations (induced by the proximity effect with a singlet superconductor), and can only modify the proximity induced triplet correlations in the presence of an exchange field $h \neq 0$ \cite{bergeret13,bergeret14}.

\begin{figure}[!tbh]
 % \centerline{\includegraphics[clip=true,width=3.5in]{fig4.eps}}
            %\centerline{\includegraphics[clip=true,width=2.5in]{fig1b.eps}}
   \begin{minipage}[b]{\linewidth}
     \centerline{\includegraphics[clip=true,width=2.4in]{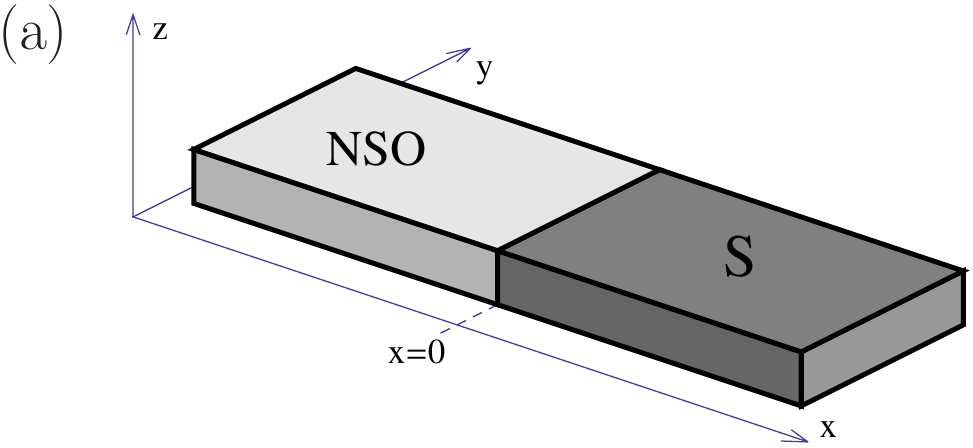}}
     \end{minipage}\hfill
    \begin{minipage}[b]{\linewidth}
   \centerline{\includegraphics[clip=true,width=2.4in]{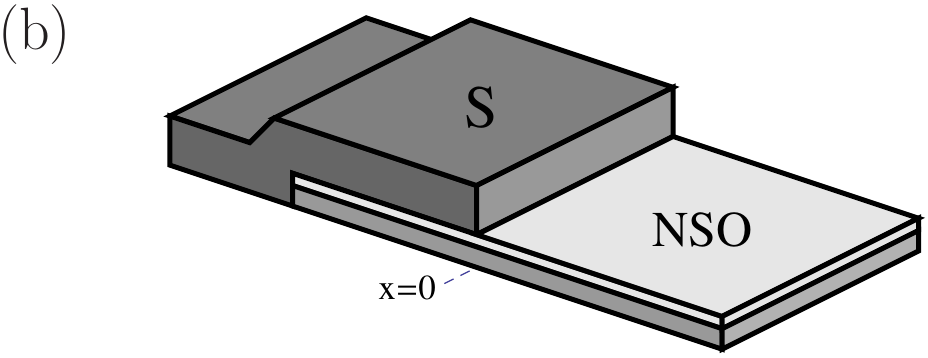}}
  \end{minipage}
   \caption{(a) Sketch of the 2D NSO/S interface under consideration. The both materials are 2D, the spin-orbit coupling $\alpha \neq 0$ only to the left of the interface, while $\alpha = 0$ to the right and there is also an intrinsic singlet superconducting pairing there. (b) Another possible realization of the proximity effect between NSO and a superconductor. The difference from panel (a) is that the interface is between the different parts of the same 2D material, therefore the value of $\alpha$ is the same for the both sides of the interface. NSO by itself is not superconducting, but the superconductivity is realized in the left part by the proximity to the singlet superconductor, which is placed on top of the NSO material.}
\label{figure}
\end{figure}

The sketch of the system is shown in Fig.~\ref{figure}(a). The interface between the superconductor and the NSO is at $x=0$ [see Fig.~\ref{figure}(a)]. The SO coupling is nonzero only in the normal metal part and is absent in the superconductor. The NSO/S interface is assumed to be fully transparent. However, we have also considered another system, where the superconducting and normal regions have absolutely the same normal state hamiltonians with non-zero spin-orbit interaction term. The corresponding experimental setup could be realized on the basis of a proximity induced superconductivity [see Fig.~\ref{figure}(b)]. We have found that the results for the proximity induced triplet correlations in NSO region are the same for the both setups.

Here we present the detailed calculations only for the case shown in Fig.~\ref{figure}(a). Our calculations are based on Eq.~(\ref{eq_g_final}). For simplicity we have considered only the linearized case here, when the Eilenberger equations can be linearized with respect to the anomalous Green's function. Under our conditions it can be realized at $T \to T_c$, where $T_c$ is a critical temperature of the superconductor.  The Green's function $\check g$ in the Nambu space can be represented as
\begin{eqnarray}
\check g = \left(
\begin{array}{cc}
\hat g & \hat f \\
\hat {\tilde f} & \hat {\tilde g}\\
\end{array}
\right),
\label{g_nambu}
\end{eqnarray}
where it is enough to calculate the normal $\hat g, \hat {\tilde g}$ components for $\Delta=0$ and the anomalous components $\hat f, \hat {\tilde f}$ of the retarded Green's function can be found from the following linear equations
\begin{eqnarray}
i v_{F,x}\partial_x f^0 + 2 \varepsilon f^0 + \Delta \tilde g^0 -\Delta g^0 = 0 ~~~~~~ \label{f_0}\\
i v_{F,x}\partial_x f^x + 2 \varepsilon f^x + \Delta \tilde g^x -\Delta g^x + i \alpha p_{F,x} f^z -~~ \nonumber \\
\frac{i\alpha p_y}{2 p_{F,x}}\partial_x f^q = 0 ~~~~~~~~~~ \label{f_x}\\
i v_{F,x}\partial_x f^y + 2 \varepsilon f^y + \Delta \tilde g^y -\Delta g^y + i \alpha p_y f^z = 0 ~~~~~~\\
i v_{F,x}\partial_x f^z + 2 \varepsilon f^z + \Delta \tilde g^z -\Delta g^z - i \alpha (p_{F,x} f^x + p_y f^y ) + \nonumber \\
\frac{i \alpha^2 p_y}{2 v_{F,x}}f^q=0, ~~~~~~~~~~~ \label{f_z}
\end{eqnarray}
where we introduce the following expansion of the anomalous Green's function $\hat f$ over the spin basis: $\hat f=f^0 \hat \sigma_0 + f^i \hat \sigma_i$.
While $f^0$ is the singlet component of the anomalous Green's function, $f^i$ for $i=x,y,z$ are the corresponding triplet components. The last terms in Eqs.~(\ref{f_x}) and (\ref{f_z}) are the corrections of the order of $\Delta_{so}/\varepsilon_F$ to the quasiclassical approximation, therefore one can use the quasiclassical approximation for the anomalous Green's function in these terms. As it was mentioned above, $\hat f^q$ has no triplet components in the absence of the exchange field, that is $\hat f^q =f^q \hat \sigma_0$.

$f^q$ can be easily found making use of the quasiclassical version of Eqs.~(\ref{f_0})-(\ref{f_z}), boundary conditions, which are reduced to continuity of the anomalous Green's function in the quasiclassical limit, and the asymptotic conditions, which require the anomalous Green's function to be non-growing functions at $x \to \pm \infty$. The resulting expressions take the form:
in the NSO:
\begin{eqnarray}
f_+^q=0, \nonumber \\
f_-^q=\frac{\Delta}{\varepsilon}e^{\frac{2i\varepsilon x}{v_{F,x}}}
\label{f_q_NSO}
\end{eqnarray}
and in the superconductor:
\begin{eqnarray}
f_+^{q,S}=\frac{\Delta}{\varepsilon}(1-e^{\frac{2i\varepsilon x}{v_{F,x}}}), \nonumber \\
f_-^{q,S}=\frac{\Delta}{\varepsilon},
\label{f_q_S}
\end{eqnarray}
where subscripts $+$ and $-$ correspond to right-moving ($v_{F,x}>0$) and left-moving ($v_{F,x}<0$) trajectories, respectively. The exponential factors in the above expressions decay at the appropriate infinity due to the fact that for the retarded Green's functions $\varepsilon$ has an infinitesimal imaginary value $i\delta$ with $\delta>0$.

In order to find the corrections of the order of $\Delta_{so}/\varepsilon_F$ to this quasiclassical solution, we need the normal components $\hat g, \hat {\tilde g}$ of the Green's function up to the same order of magnitude. It is easy to check that the following solution in the NSO region
\begin{eqnarray}
\hat g_+=-\hat {\tilde g}_- = 1, \nonumber \\
\hat g_-=-\hat {\tilde g}_+=1+\frac{\alpha p_y}{v_{F,x}p_{F,x}}\hat \sigma_x
\label{g_correction}
\end{eqnarray}
satisfies the Eilenberger equations (\ref{eq_g_final}), the normalization conditions (\ref{norm_final}) and the boundary conditions (\ref{boundary_final}).

Substituting Eq.~(\ref{g_correction}) into Eqs.~(\ref{f_0})-(\ref{f_z}) and making use of boundary conditions (\ref{boundary_final}) one obtains the following expression for the proximity induced anomalous retarded Green's function in the NSO region:
\begin{eqnarray}
f_-^0=\frac{\Delta}{\varepsilon}e^{\frac{2i\varepsilon x}{v_{F,x}}}, \label{f_0_sol} \\
f_-^x=\biggl( \frac{i \alpha p_y^3 \Delta x}{v_{F,x}^2 p_{F,x}p_F^2}+\frac{p_y \alpha \Delta}{2 v_{F,x}p_{F,x}\varepsilon} \biggr)e^{\frac{2i\varepsilon x}{v_{F,x}}}, \label{f_x_sol} \\
f_-^y=-\frac{i \alpha p_y^2 \Delta x}{v_{F,x}^2 p_F^2}e^{\frac{2i\varepsilon x}{v_{F,x}}}, \label{f_y_sol} \\
f_-^z=0,
\end{eqnarray}
while $\hat f_+ = 0$. In the superconductor the solution has no corrections to the quasiclassical answer, if the spin-orbit coupling is zero there.

It is seen from Eqs.~(\ref{f_x_sol})-(\ref{f_y_sol}) that the proximity-induced superconducting condensate in the Rashba metal has triplet components of the first order with respect to $\Delta_{so}/\varepsilon_F$ in the absence of a Zeeman term. Our answer fully coincides with the proper expansion to the first order with respect to $\Delta_{so}/\varepsilon_F$ of the general result for the Gor'kov Green's function, obtained in Ref.~\onlinecite{reeg15}, what is a good check of the validity of our approach. It is worth to mention here that expressions (\ref{f_x_sol})-(\ref{f_y_sol}) are only valid at the distances $x<\xi_s/(\Delta_{so}/\varepsilon_F)$ from the interface (where $\xi_s$ is a superconducting coherence length), because physically our approximation can be viewed as a projection of two different quasiparticle trajectories, corresponding to two different spin-orbit split Fermi surfaces, onto the same direction, determined by the Fermi surface in the absence of the spin-orbit splitting. However, this restriction is of no practical importance for the problems under consideration because all the proximity-induced superconducting correlations, which are of interest for us, decay much faster, at the characteristic length scale of $\xi_s$.

Now we discuss the symmetry classification of the obtained proximity-induced correlation. Pair amplitudes
are classified into four types according to their behavior with respect to Matsubara frequency,
momentum (parity), and spin \cite{eschrig15}. Type A: spin singlet, even frequency, even parity; type B: spin singlet, odd frequency, odd parity;
type C: spin triplet, even frequency, odd parity and type D: spin triplet, odd frequency, even parity.

In order to analyze which types of correlations are present in Eqs.~(\ref{f_0_sol})-(\ref{f_y_sol}), we should turn to the Marsubara frequency representation and divide the correlations into symmetric and antisymmetric parts with respect to $\bm p_F \to -\bm p_F$.
As for singlet correlations, here we have the both types of them. The type A correlations are the most typical and survive for a dirty case as well. The singlet, odd frequency and odd parity correlations also arise here due to the broken translational invariance, as it was reported for other physical systems with broken translational symmetry \cite{tanaka07,eschrig07}. But this type of correlations would disappear in the dirty system after averaging over trajectories due to its odd-parity nature.

As for the triplet correlations, the both possible types are also present here. It is worth to underline that the singlet-triplet mixing, reported for the homogeneous superconductor with SOC \cite{gorkov01,alicea10}, is only p-wave, that is of type C. In the homogeneous case the type D of correlations was reported in the presence of a Zeeman term or the applied supercurrent \cite{malshukov08,konschelle15}.  In spatially inhomogeneous systems the odd-frequency even-parity triplet correlations also arise due to the broken translational symmetry.

It is seen from Eqs.~(\ref{f_x_sol})-(\ref{f_y_sol}) that both $f^x$ and $f^y$ components of the triplet correlations contain as type C so as type D correlations. But after averaging over trajectories $f^x$ is zero, and $f^y$ does not disappear. It is stable against disorder and it is this triplet component that gives rise to the direct magneto-electric effect, discussed in the next section.

\section{direct magnetoelectric effect in a S/NSO/S ballistic junction}

\label{direct}

In this section we predict that the ballistic S/NSO/S Josephson junction
responses to a dc supercurrent flowing across the junction by
developing a stationary spin density oriented along the junction
interfaces. This phenomenon can be viewed as a direct magnetoelectric effect, i.e.
the Edelstein effect.
The analogous effect also takes place in normal intrinsic spin orbit coupled metals, where it was first theoretically predicted in
Refs. \onlinecite{aronov89,edelstein90} and later observed
experimentally in Refs. \onlinecite{kato04,silov04}.
In normal spin-orbit coupled metals the spin polarization is produced by externally applied electric field.
The magnetoelectric polarizability was also discussed in the normal phase of topological
insulators \cite{rev1,rev2,essin09}.
Further the  magnetoelectric effect was also predicted for
bulk superconductors \cite{edelstein95,edelstein05} and diffusive superconducting heterostructures \cite{malshukov08}, where its essence is that the supercurrent gives rise to a spin polarization along the direction, determined by the particular type of the spin-orbit coupling. It is also predicted in Josephson junctions on the basis of 3D topological insulators surface states \cite{bbza16}. Therefore, it is natural that the same effect should take place in ballistic spin-orbit coupled Josephson junctions. Below we calculate it on the basis of the formalism, developed in the present work.

To uncover this phenomenon, we first evaluate the average spin
polarization:
\begin{equation}
\bm S =\frac{1}{2} \big\langle \hat \Psi^\dagger(\bm r,t)\hat {\bm \sigma} \hat \Psi (\bm r,t)  \big\rangle
\label{spin_general}.
\end{equation}
In terms of the Green function, the components of spin polarization take the
following form:
\begin{eqnarray}
S^\alpha =-\frac{i}{8}\lim \limits_{\bm r \to \bm r'}{\rm Tr}_4 \int \frac{d\varepsilon}{2\pi} \frac{d p_y }{2 \pi} \hat \sigma^\alpha \hat \tau_z \check G^K( p_y, \bm r, \bm r',\varepsilon)= \nonumber \\
-\frac{1}{16}{\rm Tr}_4 \int \frac{d\varepsilon}{2\pi} \frac{d p_y }{2 \pi} \hat \sigma^\alpha \hat \tau_z \frac{1}{|v_{F,x}|}\biggl[ \check g_+^K + \check g_-^K \biggr]
\label{spin_quasiclassical},~~~~~~
\end{eqnarray}
where $\check g_{\pm}^K$ is the Keldysh component of the quasiclassical Green's function, which can be expressed via the retarded, advanced components and the distribution function. For the equilibrium problem we consider the above expression can be rewritten as follows:
\begin{eqnarray}
S^\alpha =
-\frac{1}{8}{\rm Tr}_4 \int \frac{d\varepsilon}{4\pi^2} \frac{d p_y }{|v_{F,x}|} \hat \sigma^\alpha \hat \tau_z \tanh \frac{\varepsilon}{2T}{\rm Re}\biggl[ \check g_+^R + \check g_-^R \biggr]
\label{spin_equilibrium}.~~~~~~
\end{eqnarray}
First of all, in order to check our formalism, we recover well-known Edelstein result\cite{edelstein95} for the spin polarization induced by the supercurrent in bulk ballistic 2D superconductor in the presence of Rashba spin-orbit coupling.
The supercurrent, flowing through the homogeneous superconductor, leads to nonzero gradient of the order parameter phase $\chi(x)$ in the direction of the current. Quasiclassical solution for the anomalous Green's functions, expanded up to the first order with respect to the superconducting order parameter phase gradient $\partial_x \chi$, have only singlet component, which takes the form
\begin{eqnarray}
f_{\pm}^q = \frac{\Delta}{\varepsilon}\bigl( 1+\frac{v_{F,x}\partial_x \chi}{2\varepsilon} \bigr) \nonumber \\
{\tilde f}_{\pm}^q = -\frac{\Delta^*}{\varepsilon}\bigl( 1+\frac{v_{F,x}\partial_x \chi}{2\varepsilon} \bigr)
\label{f_bulk_q},
\end{eqnarray}
and the quasiclassical solution for the normal Green's function can be found making use of the normalization condition
\begin{eqnarray}
g_{\pm}^q = -{\tilde g}_{\pm}^q=1-\frac{1}{2}f_{\pm}^q{\tilde f}_\pm^q= \nonumber \\
1+\frac{|\Delta|^2}{2\varepsilon^2}\bigl( 1+\frac{v_{F,x}\partial_x \chi}{\varepsilon} \bigr)
\label{g_bulk_q}.
\end{eqnarray}
The triplet corrections of the order of $\Delta_{so}/\varepsilon_F$ to the anomalous Green's function  can be found from Eqs.~(\ref{f_x})-(\ref{f_z}) and take the form
\begin{eqnarray}
f_{\pm}^x = \frac{\Delta\alpha p_y }{2v_F p_F\varepsilon}\biggl( \frac{p_y^2}{p_{F,x}^2}\bigl( 1+\frac{v_{F,x}\partial_x \chi}{2 \varepsilon} \bigr)+ 1 \biggr), \label{f_x_bulk} \\
f_{\pm}^y = \frac{\Delta}{\varepsilon}\frac{\alpha p_y^2 \partial_x \chi}{4p_F^2 \varepsilon}, ~~~~~~~~~~ \label{f_y_bulk}
\end{eqnarray}
while $f_{\pm}^z$ is zero up to the considered accuracy. ${\tilde f}_\pm^\beta=-f_\pm^\beta$, where $\beta=x,y,z$. The triplet correction to the singlet quasiclassical solution for the normal Green's function (\ref{g_bulk_q}) can also be found from the normalization condition (\ref{norm_final}) and takes the form
\begin{eqnarray}
g_{\pm}^\beta = -\frac{\alpha p_y \hat \sigma_x}{2|v_{F,x}| p_{F,x}}\bigl( 1- {\rm sgn}v_{F,x} \bigr)+ \nonumber \\
\frac{\alpha p_y \hat \sigma_x}{4 v_{F,x} p_{F,x}}f_{\pm}^q {\tilde f}_{\pm}^q - \frac{1}{2}\bigl[ f_\pm^q {\tilde f}_{\pm}^\beta+ f_\pm^\beta {\tilde f}_{\pm}^q\bigr].
\label{g_bulk}
\end{eqnarray}
Substituting Eqs.~(\ref{f_bulk_q}), (\ref{f_x_bulk}) and (\ref{f_y_bulk}) into Eq.~(\ref{g_bulk}) and, in its turn, substituting the resulting expression for $g_{\pm}^a$ into Eq.~(\ref{spin_equilibrium}), one obtains the following final result for the supercurrent-induced spin polarization in the bulk of a Rashba spin-orbit coupled superconductor:
\begin{eqnarray}
S^y=\frac{\alpha m \Delta^2\partial_x \chi}{8T^2}\sum \limits_{n\geq 0} \frac{1}{\pi^3 (2n+1)^3},
\label{spin_result_bulk}
\end{eqnarray}
while $S_x=0$, that is the induced spin polarization is perpendicular to the supercurrent direction. The above expression coincides with the Edelstein result for the ballistic case and in the limit $\alpha p_F/T \gg 1$, which we consider in the present work (one should only take into account that our $\alpha$ is twice larger than one used in Ref.~\onlinecite{edelstein95}).

\begin{figure}[!tbh]
  %\centerline{\includegraphics[clip=true,width=2.4in]{fig2_0.eps}}
            %\centerline{\includegraphics[clip=true,width=2.5in]{fig1b.eps}}
   \begin{minipage}[b]{\linewidth}
     \centerline{\includegraphics[clip=true,width=2.4in]{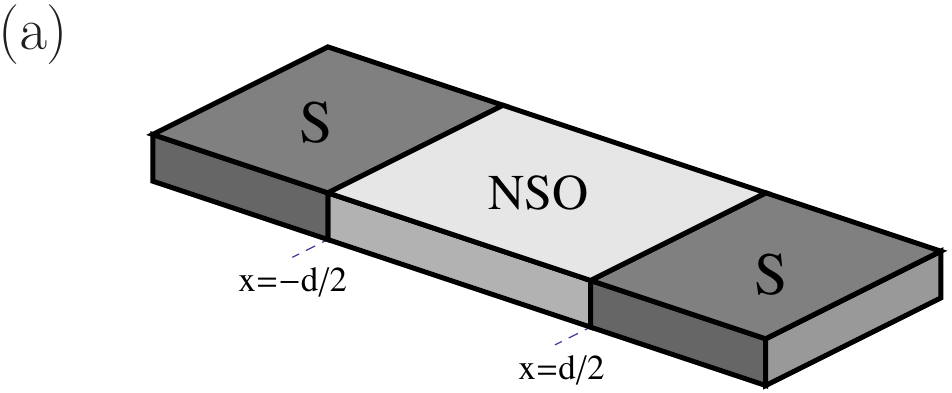}}
     \end{minipage}\hfill
    \begin{minipage}[b]{\linewidth}
   \centerline{\includegraphics[clip=true,width=2.4in]{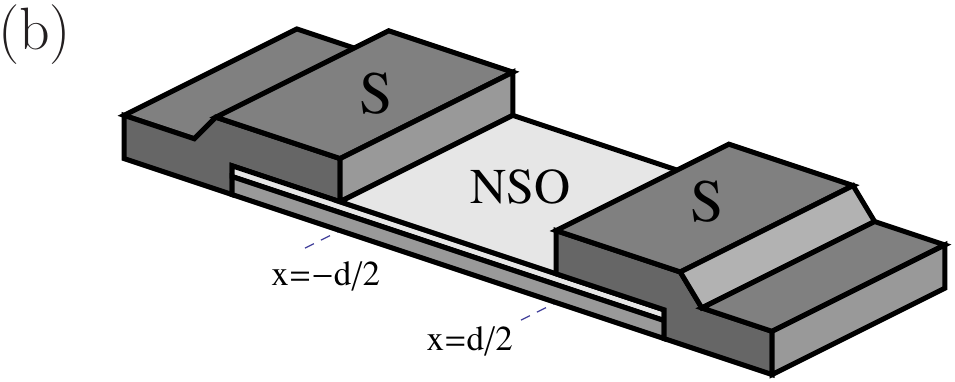}}
  \end{minipage}
   \caption{(a) Sketch of the 2D S/NSO/S junction under consideration. (b) Alternative realization of the S/NSO/S Josephson setup.}
\label{junction}
\end{figure}

Now we can turn to the case of the Josephson junction. The sketch of the system under consideration is shown in Fig.~\ref{junction}(a). The S/NSO interfaces are at $x=\mp d/2$. The difference between the setups in Figs.~\ref{junction}(a) and (b) is the same as in panels (a) and (b) of Fig.~\ref{figure}. The results for the setup in panel (b) are the same, and the calculations for this case are discussed in Appendix A.

The general scheme of the calculations is the same as for the single NSO/S interface. At first one should find the purely quasiclassical solution for the anomalous Green's functions, without the triplet corrections, which is very well known and takes the form:
\begin{eqnarray}
f_+^{q,l}=\frac{\Delta}{\varepsilon}e^{-i\chi/2} \nonumber \\
f_-^{q,l}=\frac{\Delta}{\varepsilon}e^{-i\chi/2}-\frac{2i\Delta}{\varepsilon}\sin \biggl[\frac{\varepsilon d}{v_{F,x}}-\frac{\chi}{2}\biggr] e^{\frac{2i\varepsilon x}{v_{F,x}}}
\label{f_q_l},
\end{eqnarray}
\begin{eqnarray}
f_+^{q,r}=\frac{\Delta}{\varepsilon}e^{i\chi/2}+\frac{2i\Delta}{\varepsilon}\sin \biggl[\frac{\varepsilon d}{v_{F,x}}-\frac{\chi}{2}\biggr] e^{\frac{2i\varepsilon x}{v_{F,x}}} \nonumber \\
f_-^{q,r}=\frac{\Delta}{\varepsilon}e^{i\chi/2}
\label{f_q_r},
\end{eqnarray}
\begin{eqnarray}
f_+^{q}=\frac{\Delta}{\varepsilon}e^{-i\chi/2}e^{\frac{2i\varepsilon (x+\frac{d}{2})}{v_{F,x}}} \nonumber \\
f_-^{q}=\frac{\Delta}{\varepsilon}e^{i\chi/2}e^{\frac{2i\varepsilon (x-\frac{d}{2})}{v_{F,x}}}
\label{f_q_inter},
\end{eqnarray}
where $\Delta$ is an absolute value of the superconducting order parameter, which is assumed to be the same in the both superconductors and $\chi$ - is a superconducting phase difference between the leads. Superscript $l(r)$ refers to the left (right) superconductor and the anomalous Green's function in the interlayer is defined just as $f_\pm^q$. ${\tilde f}_\pm^q=f_\mp^q(-\Delta,-\chi,-v_{F,x})$.

The normal quasiclassical functions can be found from the normalization condition and in the interlayer take the form:
\begin{eqnarray}
g_\pm^q=-{\tilde g}_\pm^q = 1+\frac{\Delta^2}{2\varepsilon^2}e^{\mp i \chi + 2i\varepsilon d/|v_{F,x}|}
\label{g_q_inter}.
\end{eqnarray}
The triplet corrections to the anomalous Green's functions are to be found from Eqs.~(\ref{f_x})-(\ref{f_z}) and boundary conditions (\ref{boundary_final}) and in the interlayer take the form:
\begin{eqnarray}
f_\pm^x=\biggl[ (x \pm \frac{d}{2})M_\pm \frac{p_y}{p_F}+ \nonumber \\
\frac{\alpha p_y \Delta}{2 v_{F,x}p_{F,x}\varepsilon}e^{\mp i \chi/2 + i\varepsilon d/|v_{F,x}|}\biggr]e^{2i\varepsilon x/v_{F,x}}
\label{f_inter_x} \\
f_\pm^y=-(x \pm \frac{d}{2})M_\pm \frac{p_{F,x}}{p_F}e^{2i\varepsilon x/v_{F,x}}
\label{f_inter_y},
\end{eqnarray}
while $f_\pm^z=0$ and
\begin{eqnarray}
M_\pm = \frac{i\alpha p_y^2 \Delta}{p_{F,x} v_{F,x}^2 p_F}e^{\mp i \chi/2 + i \varepsilon d/|v_{F,x}|}
\label{M_pm}.
\end{eqnarray}
${\tilde f}_\pm^x=f_\mp^x(-\Delta,-\chi,-v_{F,x})$, ${\tilde f}_\pm^y=-f_\mp^y(-\Delta,-\chi,-v_{F,x})$.

The triplet correction to the singlet quasiclassical solution for the normal Green's function (\ref{g_q_inter}) can be found substituting the triplet components (\ref{f_inter_x}) and (\ref{f_inter_y}) into Eq. (\ref{g_bulk}). Finally the induced spin polarization should be found from Eq.~(\ref{spin_equilibrium}) and takes the form
\begin{eqnarray}
S^y = \frac{n_2}{2}\frac{\Delta}{\varepsilon_F}\frac{\Delta}{2\pi T}\frac{\alpha p_F}{\varepsilon_F}\frac{d}{\xi}\sin \chi\times \nonumber \\
\int \limits_{-\pi/2}^{\pi/2} \frac{d\varphi}{2\pi}\tan^2 \varphi \sum \limits_{n \geq 0}\frac{e^{-\frac{d(2n+1)}{\xi \cos \varphi}}}{2n+1}
\label{spin_heterostructure},
\end{eqnarray}
while other components of the spin polarization are zero. Here $n_2=p_F^2/(2\pi)$ is the particle density, $\xi=v_F/(2\pi T)$ is the superconducting coherence length in the normal metal for a ballistic case. Integration is over angle $\varphi$ between the quasiparticle momentum and the normal to the interface.

As it was indicated in Ref.~\onlinecite{edelstein05}, the induced spin polarization in the homogeneous spin-orbit coupled superconductor Eq.(\ref{spin_result_bulk}) can be represented as
\begin{eqnarray}
\bm S = \kappa \bigl[ \bm c \times \frac{\bm j_s}{e v_F} \bigr]
\label{spin_phys},
\end{eqnarray}
where $j_s$ is the supercurrent. For the case $\alpha p_F/2 \pi T = \Delta_{so}/2\pi T \gg 1$
\begin{eqnarray}
\kappa = \frac{\alpha p_F}{8\varepsilon_F}
\label{d1}.
\end{eqnarray}
In the considered here case of S/NSO/S heterostructure Eq.~(\ref{spin_phys}) is also valid, but now $j_s$ is the Josephson current flowing through the junction. The Josephson current through the junction takes the form
\begin{eqnarray}
j_s=\frac{2 \Delta^2 p_F}{\pi^2 T}\int \limits_{-\pi/2}^{\pi/2} \frac{d\varphi}{2\pi} \cos \varphi \sum \limits_{n \geq 0}\frac{e^{-\frac{d(2n+1)}{\xi \cos \varphi}}}{(2n+1)^2}.
\label{josephson}
\end{eqnarray}
Taking into account Eq.~(\ref{josephson}) and comparing Eqs.~(\ref{spin_heterostructure}) and (\ref{spin_phys}), for $\kappa$ one can obtain
\begin{eqnarray}
\kappa =  \frac{1}{8}\frac{\alpha p_F}{\varepsilon_F}\frac{d}{\xi}\frac{\int \limits_{-\pi/2}^{\pi/2} \frac{d\varphi}{2\pi}\tan^2 \varphi \sum \limits_{n \geq 0}\frac{e^{-\frac{d(2n+1)}{\xi \cos \varphi}}}{2n+1}}{\int \limits_{-\pi/2}^{\pi/2} \frac{d\varphi}{2\pi} \cos \varphi \sum \limits_{n \geq 0}\frac{e^{-\frac{d(2n+1)}{\xi \cos \varphi}}}{(2n+1)^2}}=
\frac{\alpha p_F}{8\varepsilon_F}.~~~~
\label{d2}
\end{eqnarray}
That is $\kappa$ has the same value in different ballistic systems, such as homogeneous superconductors and transparent Josephson junctions. It would be interesting to find out if this universal behavior is valid for tunnel S/NSO/S junctions as well. It is worth to note here that the direct magneto-electric effect should also take place in S/NSO/S junctions with very strong SO coupling in the interlayer ($\Delta_{so} \sim \varepsilon_F$), but our theory is not able to describe this case quantitatively. This problem can be solved on the basis of the different quasiclassical formalism, where the SO interaction is so strong, that the coupling between the two helical subbands is disregarded \cite{agterberg07,houzet15}.

Eq.~(\ref{spin_phys}) is also valid in the quasistationary limit of the ac Josephson regime, when the Josephson current is $\propto \sin 2eVt$. In this case the induced spin polarization oscillates with the Josephson frequency.
From Eqs.~(\ref{d1}) and (\ref{d2}) it is seen that in the both cases the direct magnetoelectric effect is of the first order in $\Delta_{so}/\varepsilon_F$, the same as for the value of the induced triplet correlations, and, therefore, is absent in the quasiclassical approximation, but can be successfully described by our theory.

\section{Conclusions}

\label{conclusions}

The quasiclassical theory in terms of the equations for the Green's functions (Eilenberger equations) is generalized in order to be able to calculate the Green's functions up to the first order with respect to the parameter $\Delta_{so}/\varepsilon_F$. The Eilenberger equations are supplied by the corresponding normalization condition and the boundary conditions. It is shown that taking into account the corrections of the first order with respect to the parameter $\Delta_{so}/\varepsilon_F$ substantially modifies the normalization and boundary conditions.

The developed theory allows for quantitative description of the magneto-electric effects and proximity-induced triplet correlation in the presence of SOC as in homogeneous superconductors, so as in hybrid superconducting systems under the condition of not very large spin-orbit coupling: $\Delta \ll \Delta_{so} \ll \varepsilon_F$.

On the basis of our formalism we have considered the proximity effect at the interface between the the singlet superconductor and the Rashba metal in the ballistic limit. It is shown that the proximity-induced triplet correlations in the spin-orbit coupled normal metal are induced without any exchange or Zeeman term and their value is of the order of $\Delta_{so}/\varepsilon_F$. These correlations are long-ranged, that is they decay on the length scale of the normal state coherence length in the NSO region. They also contain an odd-frequency even-momentum component, which does not disappear after averaging over trajectories. These correlations are beyond the accuracy of the standard quasiclassical approximation, but can be described by our theory. Our result coincides with the proper expansion (in powers of $\Delta_{so}/\varepsilon_F$) of the result of Ref.~\onlinecite{reeg15} for proximity-induced triplet correlations, obtained in the framework of the exact Gor'kov technique.

The direct magneto-electric effect in superconductor/Rashba metal/superconductor ballistic junction is also considered. The quantitative result for the spin polarization, induced by the Josephson current flowing through the junction, is obtained. It is shown that by the order of magnitude the result is similar to the case of homogeneous spin-orbit coupled superconductor, but the particular value of the induced polarization depends on the length of interlayer. At the same time the ratio of the induced polarization value to the supercurrent, flowing through the junction, is exactly the same as in the homogeneous case.

In the present work we have focused on the magneto-electric effects taking place in the absence of the ferromagnetic elements or an applied magnetic field in the system. The developed theory can also be applied to the SO coupled hybrid superconducting structures in the presence of the exchange field/Zeeman term, for example, for quantitative description of the inverse magneto-electric effect and $\varphi_0$-junction behavior.

\appendix

\section{Direct magneto-electric effect for the case of proximity-induced superconducting leads}

Here the scheme of calculation of the supercurrent-induced spin polarization is presented for the system, shown in Fig.~\ref{junction}(b). The difference of this case from the one shown in Fig.~\ref{junction}(a) and considered in Sec.~\ref{direct} is that the superconducting leads and the interlayer of the S/NSO/S junction are made of the same 2D SO-coupled material, therefore the same SO coupling is present in all the parts of the system. The superconductivity is induced by proximity to singlet superconductors, which are placed on top of the SO-coupled material.

The general scheme of the calculations is the same as for the S/NSO/S interface, considered in Sec.~\ref{direct}. At first we find the purely quasiclassical solution for the anomalous Green's functions, without the triplet corrections. The SO coupling does not influence the purely quasiclassical solution, consequently it is the same and is expressed by Eqs.~(\ref{f_q_l})-(\ref{f_q_inter}) and (\ref{g_q_inter}).

The triplet corrections to the anomalous Green's functions are to be found from Eqs.~(\ref{f_x})-(\ref{f_z}), but now we should not use boundary conditions (\ref{boundary_final}). This is because they are derived for the case when SO coupling is nonzero only at one side of the interface.
Instead the appropriate boundary conditions reduce to the continuity of the Green's functions at $x=\pm d/2$ here. Solving the simple algebraic problem  we find that in this case the triplet components of the anomalous Green's function in the interlayer are the same as in Eqs.~(\ref{f_inter_x}) and (\ref{f_inter_y}), while the anomalous Green's functions in the superconducting leads differ from their values for the case considered in the main text. The reason for this difference is that now there is a nonzero SO coupling in the superconductors. Because we are only interested in the spin polarization in the interlayer, we only need the anomalous Green's functions there and, consequently, we obtain the same value of the spin polarization in this case, as in Eq.~(\ref{spin_heterostructure}).

\end{document}